# A Note on UK Covid19 death rates by religion: which groups are most 'at risk'?

Norman Fenton[1]

Queen Mary University London and Agena Ltd

10 July 2020

There has been great concern in the UK that people from the BAME (Black And Minority Ethnic) community have a far higher risk of dying from Covid19 than those of other ethnicities. However, the overall fatalities data from the Government's ONS (Office of National Statistics) most recent report on deaths by religion shows that Jews (very few of whom are classified as BAME) have a much higher risk than those of religions (Hindu, Sikh, Muslim) with predominantly BAME people. This apparently contradictory result is, according to the ONS statistical analysis, implicitly explained by age as the report claims that, when 'adjusted for age' Muslims have the highest fatality risk. However, the report fails to provide the raw data to support this. There are many factors other than just age that must be incorporated into any analysis of the observed data before making definitive conclusions about risk based on religion/ethnicity. We propose the need for a causal model for this. If we discount unknown genetic factors, then religion and ethnicity have NO impact at all on a person's Covid19 death risk once we know their age, underlying medical conditions, work/living conditions, and extent of social distancing.

Recent reports have claimed that black people in the UK have a more than four times greater risk of dying from Covid19 than whites (Booth & Barr, 2020) and that – more generally - people from the BAME (Black And Minority Ethnic) community are at much greater risk than those of other ethnicities (Apea et al., 2020).

The (Booth & Barr, 2020) article was based on a report from the UK Office of National Statistics (ONS) (Office for National Statistics, 2020a) on Covid19 deaths by *ethnic groups* (covering the period 2 March – 10 April 2020). However, a more recent ONS report on Covid19 deaths by *religion* (covering the period 2 March - 15 May) (Office for National Statistics, 2020b) provides overall numbers of fatalities for each religious group that seem to contradict the earlier report. The totals are shown here in Table 1. This table actually combines two tables from the ONS report (one for the total deaths per religious group and one for the population proportion per religion based on the most recent census).

From our Table 1 we compute in Table 2 the simple overall fatality rate for each religious group (in *deaths per 100,000*) based on the UK population size of 65 million. Curiously, the ONS report fails to provide this important summary. It shows that, based only on these population totals, Jews (by far) and then Christians have the highest death rate with atheists (no religion)[2] by far the lowest.

Table 1: Total deaths by religious group

| Religion | Covid19 deaths | Population proportion |
|---|---|---|
| No religion | 3,595 | 26.0 |
| Christian | 28,888 | 58.6 |
| Buddhist | 98 | 0.4 |
| Hindu | 594 | 1.5 |
| Jewish | 453 | 0.5 |
| Muslim | 1,307 | 4.9 |
| Sikh | 258 | 0.8 |
| Other religion | 98 | 0.4 |
| Not stated or required | 2,665 | 7.0 |

Table 2: Deaths per 100,000 by religious group

| Religion | Deaths per 100,000 |
|---|---|
| Jewish | 139 |
| Christian | 76 |
| Hindu | 61 |
| Sikh | 50 |
| Muslim | 41 |
| Buddhist | 38 |
| Other religion | 38 |
| No religion | 21 |
| Not stated or required | 59 |

---

[1] n.fenton@qmul.ac.uk, Twitter: @ProfNFenton

Now, while there are many Black and Asian Christians who come under the "BAME" (Black And Minority Ethnic) classification, almost all Jews in the UK are classified 'white' in the ONS data. So the results here seem to contradict the previous ONS report that claimed BAME were 'by far' the highest at risk group; it appears that Jews – as a distinct ethnic group – are at greatest risk.

The question is whether an obvious confounding factor like *age* is causing a Simpson's paradox effect (N. Fenton, Neil, & Constantinou, 2019; Pearl & Mackenzie, 2018). In such cases (such as shown in the sidebar for a simplified example with hypothetical data) the overall rate is higher for Group A than Group B – even though in each age sub-category the rate is higher for Group B. Indeed the sidebar refers to a real example of the paradox for US Covid19 statistics where age is a confounding factor for the higher overall fatality rate for whites.

So, is that also what we have here, i.e. is the apparently contradictory UK data simply explained by the fact that Jews and Christians are older?

This does indeed seem to be the implicit argument according in the statistical analysis in the ONS report. The report uses 'age standardized mortality rates' to take account of the age distribution differences; it concludes that Muslims, rather than Jews, have the highest fatality risk among all religious groups.

However, the report does not provide the raw data to check these 'age standardized results' (we need to know, for each age category, the number of deaths per religious group and the population proportion for each religious group in that age category) - just as the Barts study (Apea et al., 2020) failed to provide the necessary raw data to check if its bold claims about higher risk for BAME people were valid. Another concerning aspect of the report is that a lot of it focuses on the under 65s. Yet the total number of fatalities in the under 65s is dwarfed by the number of fatalities in the over 65s.

Our approach to this problem is to construct causal (probabilistic) models (N. Fenton, 2020; N. E. Fenton, Neil,

## Simpson's Paradox: hypothetical example

Suppose we have fatality data on an equal number of Blacks and Whites:

| Race | Black | White |
|---|---|---|
| *Fatalities* | | |
| No | 240 | 200 |
| Yes | 160 | 200 |
| *Fatality rate* | *40%* | *50%* |

The overall data shows that the fatality rate for Whites is higher than for Blacks.

However, when we drill down by different age groups the results are reversed:

| Age | <=65 | | >65 | |
|---|---|---|---|---|
| Race | Black | White | Black | White |
| *Fatalities* | | | | |
| No | 210 | 80 | 30 | 120 |
| Yes | 90 | 20 | 70 | 180 |
| *Fatality rate* | *30%* | *20%* | *70%* | *60%* |

In each age sub-category, the fatality rate for Blacks is higher than for Whites.

The paradox is explained by the fact that a much higher proportion of the White group here are over 65 than the proportion of over 65s in the Black group

## Simpson's Paradox in US Covid19 statistics

A real example of Simpson's paradox for Covid19 data is demonstrated by Dana Mackenzie (Mackenzie, 2020):

> ".. although in every age category (except ages 0-4), whites have a lower case fatality rate than non-whites, when we aggregate all of the ages, whites have a higher fatality rate. The reason is simple: whites are older."

---

[2] It is fair to assume these are atheists because these are people who declared "no religion" as opposed to those who did not declare any religion (i.e. those who fall into the category "not stated or required")

Osman, & McLachlan, 2020; Neil, Fenton, Osman, & McLachlan, 2020). An appropriate causal model for understanding the impact of religion and ethnicity on risk of death from Covid19 is shown in Figure 1. This is, of course, also the approach recommended by (Pearl & Mackenzie, 2018) in their excellent "Book of Why".

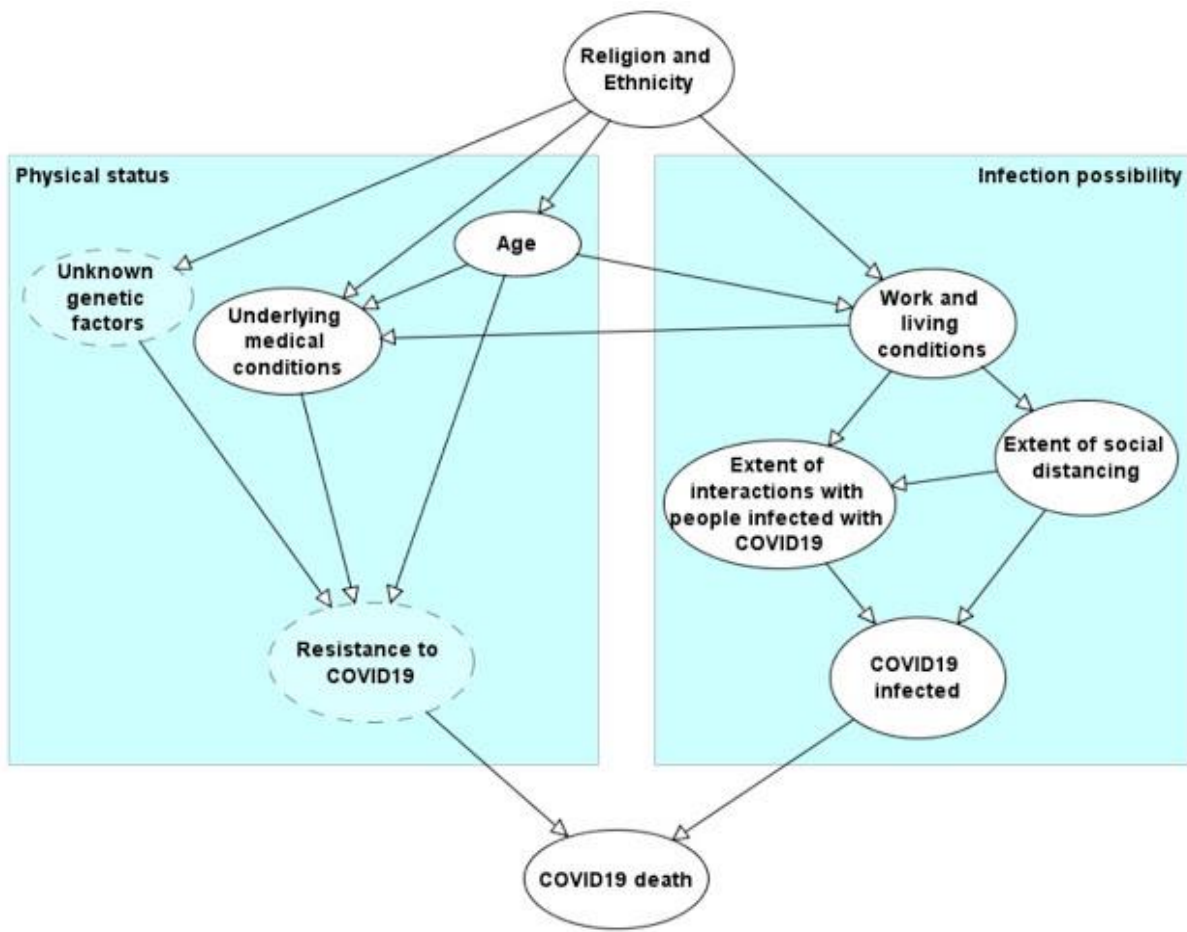

*Figure 1: The kind of causal model required to fully understand impact of religion and ethnicity on Covid19 death risk (dotted nodes represent variables that cannot be directly observed)*

Note that there are many factors other than just age that must be incorporated into any analysis of the observed data before making definitive conclusions about risk based on religion/ethnicity. Some of these factors are considered by the ONS, but without accounting for the causal interdependencies necessary when interpreting observational data. What the model shows it that, if we discount genuinely unknown genetic factors, then religion and ethnicity have NO impact at all on an individual's Covid19 death risk once we know the following for that individual: age, underlying medical conditions, work/living conditions, and extent of social distancing.

***Acknowledgement:*** Thanks to Georgina Prodhan for alerting us to the ONS report. This work was supported in part by EPSRC under project EP/P009964/1: PAMBAYESIAN: Patient Managed decision-support using Bayesian Networks.